\documentclass{Interspeech}

\interspeechcameraready

\title{Efficient Speech Enhancement via Embeddings from Pre-trained Generative Audioencoders}

\author{Xingwei}{Sun}
\author{Heinrich}{Dinkel}
\author{Yadong}{Niu}
\author{Linzhang}{Wang}
\author{Junbo}{Zhang}
\author{Jian}{Luan}

\affiliation[nocounter]{MiLM Plus}{Xiaomi Inc.}{China}
\email{\{sunxingwei, dinkelheinrich, niuyadong, wanglinzhang, zhangjunbo1, luanjian\}@xiaomi.com}

\keywords{speech enhancement, pre-trained audioencoder, audio embedding}

\usepackage{multirow}

\usepackage{cleveref}
\begin{document}

\maketitle

\begin{abstract}
Recent research has delved into speech enhancement (SE) approaches that leverage audio embeddings from pre-trained models, diverging from time-frequency masking or signal prediction techniques. This paper introduces an efficient and extensible SE method. Our approach involves initially extracting audio embeddings from noisy speech using a pre-trained audioencoder, which are then denoised by a compact encoder network. Subsequently, a vocoder synthesizes the clean speech from denoised embeddings. An ablation study substantiates the parameter efficiency of the denoise encoder with a pre-trained audioencoder and vocoder. Experimental results on both speech enhancement and speaker fidelity demonstrate that our generative audioencoder-based SE system outperforms models utilizing discriminative audioencoders. Furthermore, subjective listening tests validate that our proposed system surpasses an existing state-of-the-art SE model in terms of perceptual quality.

\end{abstract}

\section{Introduction}
Speech enhancement (SE) is a fundamental task in audio signal processing, aiming to improve the quality and intelligibility of speech signals that are degraded by noise. Most deep neural network (DNN) based SE methods primarily rely on mask or signal prediction techniques with a deep-learning framework tailored towards the SE task~\cite{Zheng2023Sixty}.  

Recent progress in self-supervised learning has given rise to robust auudioencoders pre-trained on large-scale unlabeled audio datasets, which are capable of extracting intrinsic structure and transferable representations of audio signals, such as WavLM~\cite{Chen2021WavLMLS}. The application of audio embedding from pre-trained audioencoder has been shown to be effective in various SE frameworks. For instance, Huang et al.~\cite{Huang2022InvestigatingSL} proposed using audio embedding to predict a time-frequency mask, which is then applied to the noisy speech signal. Subsequently, Hung et al.~\cite{Hung2022BoostingSE} introduced the use of cross-domain features, combining audio embedding with magnitude spectra, to enhance mask prediction further. These methods leverage the rich representations provided by audio embedding to improve mask prediction accuracy. 

In recent years, generative speech synthesis with neural vocoders has been explored in SE. This method involves denoising the acoustic features of speech signals and then using advanced neural vocoders to reconstruct the enhanced speech~\cite{Su2021HiFiGAN2SS}. Subsequently, combination of the advancement of audio embedding from pre-trained model and the generative ability of neural vocoder has been explored in SE task. In~\cite{Irvin2022SelfSupervisedLF}, a vocoder is trained to directly synthesize clean speech using paired clean and noisy speech data. In this method, the input of vocoder is the embedding of noisy speech derived from pre-trained audioencoder which remains frozen during the training process. This approach demonstrated the potential of leveraging pre-trained models to improve the SE performance. However, the frozen audioencoder may not be able to adapt to the specific characteristics of the noisy speech data, potentially limiting the overall performance. In a related line of work, Zhao et al.~\cite{Zhao2022SpeechEU} proposed generating clean speech directly from noisy embedding by combining audioencoder with vector quantization and vocoder. Song et al.~\cite{Song2024SpectrumAwareNV} further extended this approach by fusing audio embedding with transformed spectrogram as input to the vocoder, achieving better results by leveraging complementary information from multiple sources. However, these two approaches require fine-tuning all models, which can be computationally expensive and time-consuming. Additionally, the models are specifically trained for the SE task and may not be easily transferred to other tasks.

In this paper, we present a simple and extendable SE model, in which the embedding extracted from noisy speech with an audioencoder is denoised via a compact encoder network and then the clean speech is synthesized with a vocoder from the denoised embedding. In this model, the audioencoder and vocoder are both pre-trained rather than specifically trained for SE task. Furthermore, the denoise encoder can be replaced with other encoder for other tasks, such as speech dereverberation and separation. Specifically, the pre-trained audioencoder can be adapted from the open-sourced project, such as WavLM\footnote{https://github.com/microsoft/unilm/tree/master/wavlm}~\cite{Chen2021WavLMLS}, Whisper\footnote{https://github.com/openai/whisper}~\cite{Radford2022RobustSR} and Dasheng\footnote{https://github.com/XiaoMi/Dasheng}~\cite{Dinkel2024ScalingUM}. 
The vocoder is trained in self-supervised manner to synthesis the speech from the embedding with only clean speech data. The denoise encoder is trained to minimize the discrepancy between the noisy and clean embeddings. 
Throughout both training process, the pre-trained audioencoder remain frozen.  

\begin{figure*}[ht]
  \centering
  \includegraphics[width=\linewidth]{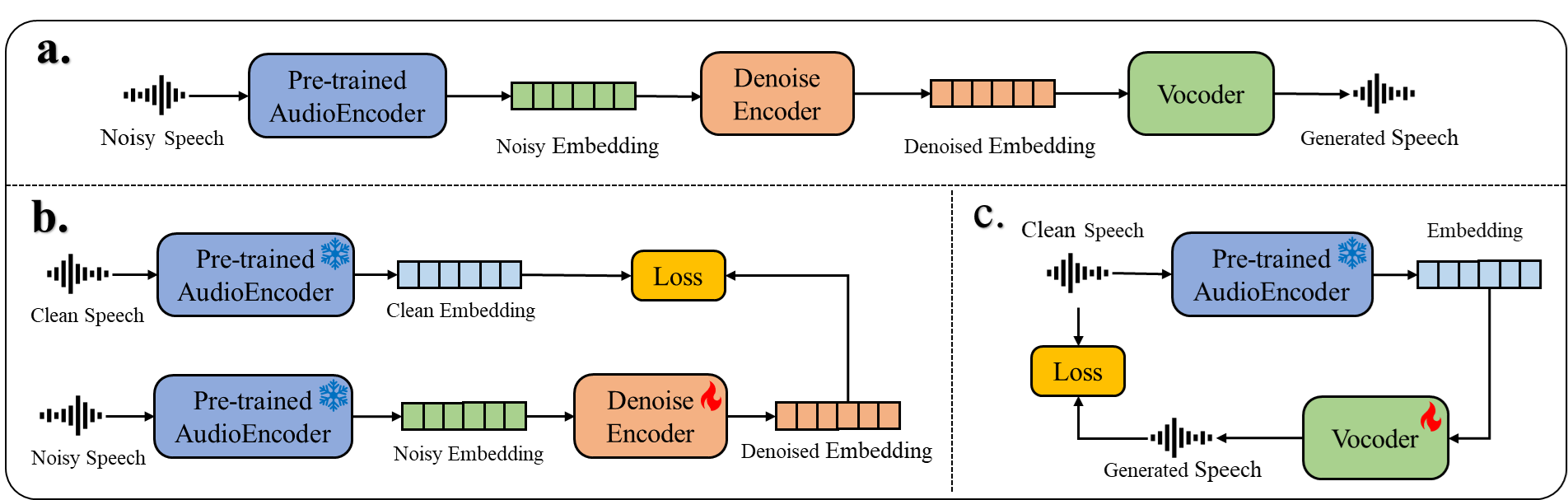}
  \caption{The proposed system: $a.$ illustrates the overall framework, which enhances speech by performing denoising at the embedding level. $b.$ depicts the training process of the proposed denoising encoder, implemented as a linear layer. $c.$ presents the training process of the vocoder.}
  \label{fig:general_framework}
\end{figure*}

In the experiments, we first compare the speech enhancement performance of our method based on different pre-trained audioencoders, including discriminative and generative models. The evaluation results demonstrate that all models achieve satisfactory improvements in noise suppression. However, the speech quality and speaker fidelity performance of denoised speech from discriminative model is significantly worse than that of the generative model. 
We also conduct ablation study on denoise encoder which indicate that our model is highly efficient, characterized by a small number of parameters and low computational requirements. It can achieve notable performance even with just two multilayer perceptrons (MLPs). Additionally, subjective listening tests indicate that our proposed system outperforms an existing state-of-the-art SE model.

\section{Proposed Approach}

As depicted in \Cref{fig:general_framework}a, our SE inference process is structured into three distinct modules. Initially, the noisy speech is input into a pre-trained audioencoder, yielding a noisy embedding. Subsequently, the denoised embedding is extracted from the noisy embedding via a denoise encoder. Ultimately, the denoised embedding is employed to produce the denoised speech by a vocoder.

The denoise encoder and vocoder are trained independently, yet both utilize the same pre-trained audio encoder. \Cref{fig:general_framework}b illustrates the training procedure for the denoise encoder. The denoise encoder is trained using Mean Squared Error (MSE) loss function between the clean and noisy embeddings, which are concurrently generated in parallel by the pre-trained audioencoder from paired noisy and clean speech samples. We simulate the noisy speech by mixing noise with the clean speech. The architecture of the denoise encoder is based on the widely-used Vision Transformer (ViT)~\cite{dosovitskiy2020vit}, employing pre-normalization and GeLU activation functions.

As depicted in \Cref{fig:general_framework}c, the vocoder is trained to synthesize speech from the embeddings provided by the pre-trained audio encoder, using only clean speech data in a self-supervised manner. 
We adopt the setup of Vocos~\cite{siuzdak2023vocos} as the vocoder, with minor adjustments to accommodate different audioencoders. 
This vocoder is capable of directly generating Fourier spectral coefficients from the speech embedding and subsequently recovering the speech waveform via inverse short-time Fourier transform. The vocoder is trained using a Generative Adversarial Network (GAN) framework, with ConvNeXt \cite{Liu2022ACF} serving as the generator backbone, and the multi-period discriminator (MPD) \cite{Kong2020HiFiGANGA} and multi-resolution discriminator (MRD)~\cite{Jang2021UnivNetAN} as discriminators. 
In addition to the adversarial loss, reconstruction loss and feature matching loss are also incorporated, as detailed in the original paper.

In this system, the pre-trained audioencoder corresponds to the encoder module of an open-source model, utilizing the checkpoint provided by the original project. Throughout the training process of both the denoise encoder and vocoder, the pre-trained audio encoders are kept frozen.

\section{Experiments}

\subsection{Data}

The training of our system leverages the clean speech and noise dataset from the ICASSP 2022 Deep Noise Suppression (DNS4) Challenge~\cite{dubey2022icassp}, with additional clean speech data from the Common Voice project~\cite{commonvoice2020} to augment the training corpus. In the vocoder training process, we exclusively employ clean speech to train the model in a self-supervised fashion. This approach enables the vocoder to learn to generate high-quality speech from embeddings without the need for paired noisy speech data. During the denoising encoder training, we dynamically mix speech and noise on-the-fly at various Signal-to-Noise Ratio (SNR) levels. The SNR level is randomly sampled from a uniform distribution ranging between -10 and 25 dB. This technique simulates noisy conditions and ensures the that the model is robust to different noise environments.

For evaluation, we assess our system using the test set without reverberation from the Interspeech 2020 Deep Noise Suppression (DNS1) Challenge~\cite{reddy2020interspeech}. This test set comprises 150 pairs of noisy and clean speech samples, providing a benchmark for assessing the performance of our noise suppression model. We also evaluate our system on the test set of the Valentini et al.~\cite{ValentiniBotinhao2017NoisySD} dataset, which consists of 824 pairs of noisy and clean speech data at four SNR levels: 2.5, 7.5, 12.5, and 17.5 dB. 

All audio data, both for training and evaluation, were resampled to a uniform sampling rate of 16 kHz to standardize the input across different datasets and ensure consistency in the model's performance. 

\begin{table*}[ht]
\centering
\begin{tabular}{l|l|c|c|cccc|c}
\hline
\multirow{2}{*}{Dataset}& \multicolumn{1}{l|}{\multirow{2}{*}{Model}} & \multicolumn{1}{c|}{\multirow{2}{*}{PESQ}} & \multicolumn{1}{c|}{\multirow{2}{*}{STOI}} & \multicolumn{4}{c|}{DNSMOS}  & \multicolumn{1}{c}{\multirow{2}{*}{NISQAv2}} \\ \cline{5-8}
& \multicolumn{1}{l|}{} & \multicolumn{1}{c|}{}  & \multicolumn{1}{c|}{}    & \multicolumn{1}{c}{P808} & \multicolumn{1}{c}{SIG} & \multicolumn{1}{c}{BAK} & \multicolumn{1}{c|}{OVRL} & \multicolumn{1}{c}{} \\ \hline
\multirow{5}{*}{Valentini} & Noisy & 1.97  & \textbf{0.92}      & 3.09    & 3.32   & 3.11   & 2.68   & 3.04     \\
 & LMS    & 1.77  & 0.85  & 3.45    & 3.12   & 4.07   & 2.90   & 4.25   \\
 & Whisper & 1.31  & 0.80  & 3.39  & 3.47   & 4.12   & 3.22 & 3.83  \\
 & WavLM   & 1.27  & 0.81  & 3.49  & \textbf{3.49}  & 4.13 & \textbf{3.26}  & 4.12    \\
 & Dasheng  & \textbf{2.32}      & 0.90  & \textbf{3.52}   & 3.45   & \textbf{4.14}  & 3.22   & \textbf{4.51}   \\ \hline    \hline
\multirow{5}{*}{DNS1}      & Noisy & 1.58  & \textbf{0.92}     & 3.16    & 3.39   & 2.62   & 2.48   & 2.58     \\
& LMS    & 1.90  & 0.90  & 3.87    & 3.34   & 4.07   & 3.10   & 3.85     \\
& Whisper  & 1.20 & 0.81 & 3.87 & 3.60   & 4.14  & 3.36  & 4.09   \\
& WavLM   & 1.20  & 0.83  & 3.91    & \textbf{3.61}  & \textbf{4.18}  & \textbf{3.40}   & \textbf{4.41}   \\
& Dasheng  & \textbf{2.24}      & \textbf{0.92}      & \textbf{4.03}   & 3.58   & 4.17   & 3.37   & \textbf{4.41}   \\ \hline
\end{tabular}
\caption{Speech enhancement evaluation results of proposed SE system with different audioencoders, where for all metrics higher is better and best results are in bold.}
\label{tab:audioencoder}
\end{table*}

\subsection{Model Setups}

We conducted a comparative analysis of the speech enhancement performance of various pre-trained models serving as audioencoders, including WavLM, Whisper, and Dasheng. To ensure a fair comparison, we selected the base version of WavLM and Dasheng, the small version of Whisper, in which there are approximately 90 million parameters in the encoder modules. Specifically, the audioencoders utilized in our experiments are referred to as WavLM, Whisper, and Dasheng. The dimension of the embeddings derived from the last layer of these audio encoders is uniformly set to 768. Additionally, we employed the log-Mel spectrogram (LMS) with a 100-dimension as a hand-crafted embedding. It is important to note that WavLM and Whisper are considered as discriminative audioencoders, whereas Dasheng is a generative audioencoder. Both types of audioencoders rely on pre-training on large datasets and excel at robust feature extraction. However, discriminative audioencoders, such as WavLM and Whisper, are adept at identifying distinct patterns within audio inputs, which focus on distinguishing phonetic and linguistic features. In contrast, generative audioencoders, such as Dasheng, concentrate on learning the holistic features of audio signals, which allows the encoded embedding to be used for both classification and audio generation tasks.

Except for LMS, our denoise encoder uses three ViT layers (ViT3) with a 768-dimensional embedding and 8 attention heads. 
For the LMS audioencoder, we add two MLP layers: one before and one after the denoise encoder, upsampling to 768 and downsampling to 100, denoted as $\text{MLP}_{\text{ViT3}}$.

To further demonstrate the scalability of our denoise encoder, we trained four models of varying sizes and architectures combined with the Dasheng audioencoder. These models include: a single-layer ViT (ViT1); three layers of bidirectional Long Short-Term Memory (LSTM) networks (BLSTM3) and three layers of LSTMs (LSTM3), both with a hidden size of 256; and two MLP layers with a hidden size of 768 (MLP2). It is important to note that the same vocoder, trained with the Dasheng model, is employed for all five distinct denoise encoders during the SE inference process.

The parameter numbers of the network models used in this study are shown in \Cref{tab:Parameter_numbers}.

\begin{table}[ht]
\renewcommand{\arraystretch}{1.1}
\centering
\begin{tabular}{c|c||c|c||c|c}
\toprule
{AE} & {\#M}  & {DE} & {\#M}    & {VO}    & {\#M} \\ 
\midrule

LMS   & 0  &  $\text{MLP}_{\text{ViT3}}$ & \textcolor{red}{14.3}     & Vocos     & \textcolor{red}{15.6}     \\ \hline
Whisper     & \textcolor{blue}{87.0}     & \multirow{3}{*}{ViT3} & \multirow{3}{*}{\textcolor{red}{14.2}} & Vocos     & \textcolor{red}{19.0} \\ \cline{1-2} \cline{5-6} 

WavLM   & \textcolor{blue}{94.4}     & & & Vocos     & \textcolor{red}{18.9}     \\ \cline{1-2} \cline{5-6} 
\multirow{5}{*}{Dasheng} & \multirow{5}{*}{\textcolor{blue}{85.4}} & & & \multirow{5}{*}{Vocos} & \multirow{5}{*}{\textcolor{red}{19.4}} \\ \cline{3-4}

   & & ViT1     & \textcolor{red}{4.7}      &  & \\ \cline{3-4}
   & & BLSTM3    & \textcolor{red}{4.8}      &  & \\ \cline{3-4}
   & & LSTM3     & \textcolor{red}{2.0}      &  & \\ \cline{3-4}
   & & MLP2      & \textcolor{red}{1.2}      &  & \\ 
   \bottomrule
\end{tabular}
\caption{Parameter numbers of different audioencoder (AE), denoise encoder (DE), and vocoder (VO) used in this work. Each row represents one system setting. $\#M$ means millions of parameters. Parameters in \textcolor{red}{red} are fine-tuned, while \textcolor{blue}{blue} are frozen.}
\label{tab:Parameter_numbers}
\end{table}

\subsection{Evaluation Metrics}

For the assessment of noise suppression performance, we utilize the Short-Time Objective Intelligibility (STOI) \cite{Taal2011AnAF} and Perceptual Evaluation of Speech Quality (PESQ) \cite{Rix2023Perceptual}, which are intrusive metrics requiring a clean speech reference. We also explore two non-intrusive metrics: DNSMOS P.835 \cite{reddy2022dnsmos} and NISQAv2 \cite{Mittag2021NISQAAD}. DNSMOS P.835 is generated by a DNN model trained with subjective evaluation scores of noise suppression clips, which include the P.808 score indicating the overall quality of the audio clip, and the quality scores of speech and background noise in addition to the overall quality: speech quality (SIG), background noise quality (BAK), and the overall quality (OVRL). NISQAv2 scores are also generated by a deep learning-based model with more insight into the cause of quality degradation, including Noisiness, Coloration, Discontinuity, Loudness, and an overall speech quality. The overall score of NISQAv2 is reported in our results.

In addition to evaluating conventional speech quality metrics, we further assess speaker fidelity through a speaker verification framework by comparing the enhanced speech with its clean reference. Specifically, we employed the ECAPA-TDNN model \cite{Desplanques2020ECAPATDNNEC}, which has been pre-trained and integrated into the Speechbrain toolkit \cite{Ravanelli2021SpeechBrainAG}, to extract speaker embeddings from the clean and enhanced signals. The fidelity metric is then quantified by computing the cosine similarity between these two embeddings.

\begin{table*}[ht]
\centering
\begin{tabular}{l|l|c|c|cccc|c}
\toprule

\multirow{2}{*}{Dataset}   & \multicolumn{1}{l|}{\multirow{2}{*}{Model}} & \multicolumn{1}{c|}{\multirow{2}{*}{PESQ}} & \multicolumn{1}{c|}{\multirow{2}{*}{STOI}} & \multicolumn{4}{c|}{DNSMOS} & \multicolumn{1}{c}{\multirow{2}{*}{NISQAv2}} \\ \cline{5-8}
\addlinespace[1pt]
 & \multicolumn{1}{l|}{}   & \multicolumn{1}{c|}{}  & \multicolumn{1}{c|}{}  & {P808} & {SIG} & {BAK} & \multicolumn{1}{c|}{OVRL} & \multicolumn{1}{c}{}     \\ 
 \midrule
 
\multirow{5}{*}{Valentini} & ViT3 & 2.32   & 0.90 & 3.52  & 3.45  & 4.14  & 3.22  & 4.51   \\
 & ViT1 & 2.33   & 0.90 & 3.52  & 3.44  & 4.14  & 3.21  & 4.47   \\
 & BLSTM3  & 2.14   & 0.89   & 3.52  & 3.42  & 4.14  & 3.19  & 4.52   \\
 & LSTM3   & 2.10 & 0.88   & 3.50   & 3.40  & 4.13  & 3.17  & 4.54   \\
 & MLP2 & 2.10 & 0.87   & 3.45  & 3.37  & 4.11  & 3.13  & 4.36   \\ \hline \hline
\multirow{5}{*}{DNS1}      & ViT3 & 2.24   & 0.92   & 4.03  & 3.58  & 4.17  & 3.37  & 4.41   \\
 & ViT1 & 2.17   & 0.92   & 4.02  & 3.58  & 4.18  & 3.37  & 4.38   \\
 & BLSTM3  & 1.96   & 0.90 & 4.01  & 3.57  & 4.19  & 3.36  & 4.50    \\
 & LSTM3   & 1.90 & 0.90 & 3.98  & 3.55  & 4.18  & 3.34  & 4.45   \\
 & MLP2 & 1.79   & 0.88   & 3.89  & 3.50  & 4.16  & 3.28  & 4.31   \\ \hline
\end{tabular}
\caption{Speech enhancement evaluation results of proposed SE system with different denoise encoders.}
\label{tab:denoise_encoder}
\end{table*}

\section{Results}

\subsection{Evaluation of Different Audioencoders}

\Cref{tab:audioencoder} presents the evaluation results for both Valentini and the DNS1 test sets. When considering intrusive metrics such as PESQ and STOI, the performance of all methods is decreased over the noisy signal, with the exception of the PESQ scores for Dasheng and LMS in the DNS1 test set. In contrast, non-invasive metrics like DNSMOS and NISQAv2 reveal a distinct performance improvement of audioencoder based methods in both noise suppression and speech quality compared with LMS and noisy speech.

The divergent performance trends of the WavLM and Whisper models, across both intrusive and non-invasive metrics,  could stem from the fact that the embeddings produced by these discriminative audioencoders capture essential phonetic and linguistic details but fail to preserve the timbre and pitch information. Therefore, the generated clean speech from the vocoder with these embeddings achieves a high level of perceptual quality for human listeners, albeit with some alteration in speech characteristics. On the other hand, the Dasheng model exhibits a more consistent performance, suggesting that its generative audioencoder embeddings capture all speech information, with the denoise encoder effectively eliminating noise components without compromising the integrity of the speech signal.

To further compare the performance of discriminative and generative audioencoders, we performed speaker fidelity evaluation on the denoised speech. The results are presented in \Cref{tab:speaker} in which the values indicate the similarity between the speaker embeddings of clean and denoised speech extracted by ECAPA-TDNN model \cite{Desplanques2020ECAPATDNNEC}. Across both datasets, the Dasheng model demonstrated superior performance in speaker fidelity tasks. In contrast, the WavLM and Whisper models yielded particularly poor results, which supports the notion that speaker information may be lost within their embeddings.

\begin{table}[ht]
\centering
\begin{tabular}{l|c|c}
\toprule
\multirow{2}{*}{Model} &  \multicolumn{2}{c}{Dataset}  \\
   & {Valentini} & {DNS1} \\ \midrule
LMS     & 0.718     & 0.847       \\ 
WavLM     & 0.408     & 0.486       \\ 
Whisper  & 0.406     & 0.489       \\ 
Dasheng   & \textbf{0.783}     & \textbf{0.881}       \\ 
\bottomrule

\end{tabular}
\caption{Speaker fidelity evaluation results of proposed SE system with different audioencoders. Results denote speaker similarity, where higher is better and best results are in boldface.}
\label{tab:speaker}
\end{table}

\vspace{-3em}
\subsection{Denoise Encoder Ablation}

The evaluation results for various denoise encoders are summarized in \Cref{tab:denoise_encoder}. These SE systems are all based on the Dasheng audioencoder. As shown in \Cref{tab:denoise_encoder}, it is evident that as the model parameter number decreases, there is a significant decline in performance with respect to the intrusive metrics, PESQ and STOI. However, the reduction in performance for non-invasive metrics, such as DNSMOS and NISQAv2, is less pronounced, suggesting that these denoise encoders can still achieve satisfactory speech perceptual quality. Notably, the performance decrease of VIT1 and BLSTM3 models is negligible when assessed with DNSMOS and NISQAv2 metrics, indicating that these models can maintain high perceptual quality despite having fewer parameters.

\subsection{Subjective Evaluation} 

In our perceptual listening study, we compared our SE system with Demucs\footnote{https://github.com/facebookresearch/denoiser}, a renowned waveform-to-waveform SE system~\cite{Dfossez2020RealTS}. 
For comparative purposes, we also included unprocessed noisy speech, clean speech, and speech denoised using the LMS method. 
\Cref{tab:subjective_mos} presents the average Mean Opinion Scores (MOS) \cite{ITU2015Subjective} for each method, as evaluated by 17 independent listeners across 15 audio samples. 
The results indicate that all SE methods significantly improve the perceptual MOS compared to the unprocessed noisy speech. 
Notably, our proposed Dasheng model outperforms Demucs, achieving an improvement of 0.76 in MOS. Furthermore, we have open-sourced the inference code with pre-trained checkpoint and audio demonstrations\footnote{https://github.com/xiaomi-research/dasheng-denoiser}.

\begin{table}[ht]
\centering
\begin{tabular}{l|c}
\hline
Method    & \multicolumn{1}{c}{MOS} \\ \hline
Noisy   & 1.61    \\ 
LMS       & 2.98    \\ 
Demucs    & 3.11    \\ 
Dasheng       & \textbf{3.87}  \\ \hline \hline
Clean     & 4.43    \\ \hline
\end{tabular}
\caption{Results of subjective listening study.}
\label{tab:subjective_mos}
\end{table}

\vspace{-3em}
\section{Discussion and Conclusion}

Global fine-tuning of all parameters within our SE system
might yield performance improvement. However, we chose not to implement global fine-tuning due to several compelling considerations. Primarily, by utilizing a pre-trained audioencoder, we can leverage pre-computed embeddings from other tasks without incurring additional computational expenses. Moreover, the pre-trained vocoder can be repurposed for diverse audio processing tasks, such as dereverberation and bandwidth expansion.

In summary, we propose a straightforward and extensible SE system that employs pre-trained audioencoder and vocoder. Experimental results demonstrate that using a generative audioencoder achieves more robust speech enhancement performance compared to discriminative audioencoders. Furthermore, the parameter efficiency of the denoise encoder is confirmed by our ablation study. A subjective listening study validates the quality of our proposed SE system with human listeners.

\bibliographystyle{IEEEtran}
\bibliography{mybib}

\end{document}